\documentclass{iau} 
\usepackage{graphicx}

\title[X-ray properties of eight  LLAGNs] %% give here short title %%
{The sample of eight  LLAGNs: \\  X-ray properties.}

\author[Nadiia G. Pulatova,  Anatoliy V. Tugay \&  Lidiia V. Zadorozhna ]   %% give here short author list %%
{Nadiia G. Pulatova$^1$,
  Anatoliy V. Tugay$^2$
   \and Lidiia V. Zadorozhna$^2$}

\affiliation{$^1$ Main Astronomical Observatory of the National Academy of Sciences of Ukraine, \\ Akademika Zabolotnoho str. 27, Kyiv, Ukraine, 03143 \\ email: {\tt nadya@mao.kiev.ua} \\[\affilskip]
$^2$Taras Shevchenko National University of Kyiv, Physical faculty, \\ ave. Glushkova 2, building 1, Kyiv, Ukraine, 03680 \\emails: {\tt tugay.anatoliy@gmail.com,
 zadorozhna\_lida@ukr.net}}

\pubyear{2021}
\volume{xxx}  %% insert here IAU Symposium No.
\jname{IAU Symposium 367/VIRTUAL
Education and Heritage in the
Era of Big Data in Astronomy}
\editors{A.C. Editor, B.D. Editor \& C.E. Editor, eds.}
\begin{document}

\maketitle

\begin{abstract}
LLAGN are very important objects for studying as they are found in a large fraction of all massive galaxies. Nevertheless this topic needs more investigation as fraction of LLAGN in all AGN are much more higher than fraction of researches dedicated to LLAGN among all AGN studies.
The goal of our work is checking out X-ray properties of LLAGN. For this purpose we created a sample of LLAGN by selecting most prominent LLAGN from literature and analyzed their X-ray spectral properties. As a result, we obtained 12 LLAGN and for 8 of them XMM X-ray observations are available. The spectra from one XMM camera, PN, were fitted with power law + absorption of neutral hydrogen. In the current report we present the previous results of this study. We plan to increase numbers of objects in our future studies. 

%% add here a maximum of 10 keywords, to be taken form the file <Keywords.txt>
\end{abstract}

\firstsection % if your document starts with a section,
              % remove some space above using this command.
\section{Introduction}

For nowadays numerous observations in all electromagnetic spectrum are available for public using and searching for a new ideas with a purpose to clarify the nature of AGN (active galactic nuclei). Mostly all scientific attention is paid to bright AGNs: Sy1 and Sy2 types. In our current study we will investigate X-ray properties of Low-Luminosity AGN (LLAGN). It is generally accepted that most time of its “life” galaxy spend in a quiescent mode. Therefore all studies connected with LLAGN plays very important role in understanding of the nature and physical mechanisms that take place in a galactic nuclei. 
The intrinsic faintness, i.e.
$ L_{Bol}<10^{44}erg\cdot s^{-1}$
, and the low level of activity are the distinctive characteristics of LLAGNs, which mainly consists of local Seyfert galaxies and Low-Ionization Nuclear Emission-line Regions (LINERs).

\begin{table}
  \begin{center}
  \caption{The sample of selected LLAGNs $^1$.}
  \label{tab1}
 {\scriptsize
  \begin{tabular}{|l|c|c|c|c|c|c|c|c|}\hline 
 {\bf Object } & {\bf RA} & {\bf DEC} & {\bf Optical } & {\bf  z} & {\bf Morph. } & {\bf Flux, }  & {\bf  X-Ray Flux, } & {\bf XMM}\\ 
{\bf  name} & & & {\bf classif.}  & & {\bf  type}  & {\bf  V mag} & {\bf erg/cm2/s } &  {\bf obs.}\\
   \hline
{\bf NGC 404} & 01 09 27.1  & +35 43 05 & LINER &	0.000107 & dE/S0 & 11.73  & $ 6.71*10-14$, $^3$	& no \\
{\bf NGC 1052}	&	02 41 04.8	&	-08 15 20.8	&	LINER h 	&	0.004930 	&	E	&	10.47 	&	$2.7*10^{-11}$, $^2$	&	5 \\
{\bf NGC 3368}	&	10 46 45.7 	&	+11 49 11.7 	&	LINER	&	0.003012 	&	SBa 	&	9.25 	&	$2.1*10^{-13}$, $^4$	&	no \\
{\bf NGC 3486}	&	11 00 23.9 	&	+28 58 29.3 	&	LINER/Sy 2	&	0.002272 	&	SABc 	&	10.53 	&	$1.1*10^{-13}$, $^5$	&	1 \\
{\bf NGC 3642}	&	11 22 18.0 	&	+59 04 27.2 	&	LINER b 	&	0.00529	&	SAbc 	&	14.04 	&	 -	&	no \\
{\bf NGC 3998}	&	11 57 56.1 	&	+55 27 12.9 	&	LINER 1.9 	&	0.003646 	&	SA0 	&	12.10 	&	$1.4*10^{-11}$, $^2$	&	2 \\
{\bf NGC 4203}	&	12 15 05.0 	&	+33 11 50.4 	&	LINER b	&	0.00362	&	SA0 	&	11.99 	&	$31.42*10^{-13}$, $^7$	&	no \\
{\bf NGC 4552}	&	12 35 39.8 	&	+12 33 22.8 	&	LINER/Sy 2/HII	&	0.000914 	&	S0 	&	9.75 	&	$9.60*10^{-13}$, $^6$	&	1 \\
{\bf NGC 4579}	&	12 37 43.5 	&	+11 49 05.1	&	LINER 1.9/Sy 1.9 	&	0.00562 	&	Sab 	&	9.66 	&	$8.9*10^{-12}$, $^2$	&	2 \\
{\bf NGC 4594}	&	12 39 59.4 	&	-11 37 22.9 	&	Sy 2/Sy 1.9	&	0.003416	&	Sa 	&	8.00 	&	$29.24*10^{-13}$,  $^6$	&	1 \\
{\bf M 87}	&	12 30 49.4 	&	+12 23 28.0 	&	LINER	&	0.00420 	&	E-E/S0 	&	8.63 	&	$1.05*10^{-10}$ ,  $^8$	&	3 \\
{\bf M 81}	&	09 55 33.1 	&	+69 03 55.0 	&	LINER/Sy1.8/ QSO	&	-0.000140 	&	SAab 	&	6.94 	&	$3.10*10^{-11}$,  $^5$	&	5 \\

 \hline
  \end{tabular}
  }
 \end{center}
\vspace{1mm}
 \scriptsize{
 {\it Notes:}\\
  $^1$ Parameters for every galaxy were taken from NED (http://ned.ipac.caltech.edu/), SIMBAD (http://simbad.u-strasbg.fr/) and Hyperleda (http://leda.univ-lyon1.fr/) databases. \\
$^2$ 15-150keV, $^3$ 0.3-8keV, $^4$  0.7-7keV,  $^5$ 2-10 keV,  $^6$ 0.2-4 keV, $^7$  0.2-4 keV, $^8$   0.1-2.4 keV}
\end{table}

\begin{table}
  \begin{center}
  \caption{The sample of selected LLAGNs $^1$.}
  \label{tab2}
 {\scriptsize
  \begin{tabular}{|l|c|c|c|c|c|c|c|c|c|}\hline 
 {\bf Object}	&  {\bf Obs ID}	&  {\bf Obs Date}	& {\bf Dur., s} 	& {\bf XSPEC } & {\bf Pho } & {\bf $N_H, 10^{22}$}	& {\bf kT , } 	& {\bf kT , } 	& {\bf $\chi^2/d.o.f.$}	\\
 {\bf  name}	&		&		&		&	 {\bf  Model}	&	 {\bf  Index}	&	&	{\bf  (ap/mek)  keV} 	& {\bf (zgauss)  keV} 	& \\

   \hline

{\bf NGC 1052}	&	0093630101	&	2001-08-15	&	16308	&	apec+zgauss+	&	1.33	&	2.28	&	0.81	&	5.96	&	55/56	\\
&		&		&		&	 po+ phabs·po	&	$\pm 0.15$	&	$\pm 0.97$	&		&		&		\\

	&	0306230101	&	2006-01-12	&	54909	&	apec+zgauss+ 	&	1.03	&	6.42	&	0.74	&	5.65	&	336/296	\\
	&		&		&		&	po+ phabs·po	&	$\pm 0.16$	&	$\pm 1.08$	&		&		&		\\
	&	0553300301	&	2009-01-14	&	52319	&	apec+zgauss+ 	&	1.48	&	9.26	&	0.75	&	6.41	&	282/287	\\
	&		&		&		&	po+ phabs·po	&	$\pm 0.074$	&	$\pm 0.56$	&		&		&		\\
	&	0553300401	&	2009-08-12	&	59018	&	apec+zgauss+ 	&	1.13	&	9.95	&	0.75	&	6.37	&	356/324	\\
	&		&		&		&	po+ phabs·po	&	$\pm 0.067$	&	$\pm 0.51$	&		&		&		\\
	&	0790980101	&	2017-01-17	&	70500	&	apec+zgauss+ 	&	1.06	&	9.59	&	0.75	&	6.37	&	293/264	\\
	&		&		&		&	po+ phabs·po	&	$\pm 3.62$	&	$\pm 5.29$	&		&		&		\\
{\bf NGC 3486}	&	0112550101	&	2001-05-09	&	15172	&	powerlaw	&	2.9	&		&		&		&	3/3	\\
	&		&		&		&		&	$\pm 0.50$	&		&		&		&		\\
{\bf NGC 3998}	&	0790840101	&	2016-10-26	&	25000	&	phabs·po	&	1.86	&	2.13·$10^{-2}$	&		&		&	158/163	\\
	&		&		&		&		&	$\pm 0.03$	&	$\pm 0.0061$	&		&		&		\\
	&	0090020101	&	2001-05-09	&	13222	&	phabs·po	&	1.85	&	2.31·$10^{-2}$	&		&		&	403/373	\\
	&		&		&		&		&	$\pm 0.015$	&	$\pm 0.0029$	&		&		&		\\
{\bf NGC 4552}	&	0141570101	&	2003-07-10	&	44838	&	phabs·po+apec	&	2.1	&	7.59·$10^{-2}$	&	0.72	&		&	96/120	\\
	&		&		&		&		&	$\pm 0.085$	&	$\pm 0.016$	&		&		&		\\
{\bf NGC 4579}	&	0790840201	&	2016-12-06	&	23002	&	po·phabs + mekal	&	1.8	&	3.82·$10^{-2}$	&	0.39	&		&	510/ 452	\\
	&		&		&		&		&	$\pm 0.013$	&	$\pm 0.0018$	&		&		&		\\
	&	0112840101	&	2003-06-12	&	23669	&	po·phabs + mekal	&	1.8	&	2.62·$10^{-2}$	&	0.62	&		&	396/ 309	\\
	&		&		&		&		&	$\pm  0.024$	&	$\pm  0.0043$	&		&		&		\\
{\bf NGC 4594}	&	0084030101	&	2001-12-28	&	43456	&	po·phabs	&	1.9	&	0.16	&		&		&	116/111	\\
	&		&		&		&		&	$\pm 0.038$	&	$\pm 0.0098$	&		&		&		\\
{\bf M 81}	&	0112521101	&	2002-04-16	&	11518	&	phabs·po+apec	&	1.87	&	0.2	&	0.93	&		&	187/ 187	\\
	&		&		&		&		&	$\pm 0.030$	&	$\pm 0.0078$	&		&		&		\\
	&	0200980101	&	2004-09-26	&	119166	&	phabs·po+apec+ 	&	2.55	&	0.25	&	1.01	&	2.15·10-3	&	818/712	\\
	&		&		&		&	zgauss	&	$\pm 0.16$	&	$\pm 0.021$	&		&		&		\\
	&	0693851101	&	2012-11-16	&	13316	&	phabs·po+apec+	&	1.74	&	0.23	&	1.23	&	1.55	&	178/164	\\
	&		&		&		&	 zgauss	&	$\pm 0.069$	&	$\pm  0.014$	&		&		&		\\
{\bf M 87}	&	0114120101	&	2000-06-19	&	60109	&	phabs·po+apec	&	2.45	&	8.85·10-2	&	1.32	&		&	922/508	\\
	&		&		&		&		&	$\pm  0.032$	&	$\pm 0.0041$	&		&		&		\\

 \hline
  \end{tabular}
  }
 \end{center}
\vspace{1mm}
 \scriptsize{
 {\it Notes:}\\
  $^1$ Parameters for every galaxy were taken from NED (http://ned.ipac.caltech.edu/), SIMBAD (http://simbad.u-strasbg.fr/) and Hyperleda (http://leda.univ-lyon1.fr/) databases. \\
$^2$ 15-150keV, $^3$ 0.3-8keV, $^4$  0.7-7keV,  $^5$ 2-10 keV,  $^6$ 0.2-4 keV, $^7$  0.2-4 keV, $^8$   0.1-2.4 keV}
\end{table}

\section{Overview}

The common properties of LLAGNs were studied on the base of different samples of LLAGNs. Very good review of LINER`s properties, the most common form of LLAGNs, was made by \cite [Dan Maoz (2008)]{Maoz2008}. In this paper author argues that there is actually more similarity than contrast between the observed spectral properties of different AGNs, with only a mild and continuous change in the properties as a function of decreasing accretion rate, going from quasars to the faint LINERs in nearby galaxies. A sample of seven nearby, early type spiral galaxies was studied in\cite[ Roberts at all (2001)]{Roberts2001}. The authors used ROSAT HRI spatial data and ASCA spectral measurements to address the question of whether a LLAGN is present in galaxies that have a LINER 2 classification. All seven galaxies have X-ray spectra consistent with a two-component, soft thermal plus hard power-law, spectral form. In the paper an idea is claimed, that in many, perhaps the majority, of LINER 2 galaxies, the nuclear X-ray luminosity does not derive directly from the presence of a LLAGN.
 \cite[Minfeng Gu \& Xinwu Cao (2009)]{GuCao09} found an interesting difference between Low-Ionization Nuclear Emission-line Regions (LINERs) and luminous active galactic nuclei (AGNs). A significant anti-correlation was revealed between the hard X-ray photon index and the Eddington ratio$ L_{Bol}/L_{Edd}$ for a sample of LINERs and local Seyfert galaxies compiled from literature with Chandra or XMM-Newton observations.
To the most brightest LLAGNs were dedicated numerous studies by many scientific groups from different countries. For example, M87 that is included to our sample of LLAGNs. In \cite[Jolley, E.J.D., Kuncic Z. (2007)]{JolleyKuncic07} to the well known LLAGN M87 was applied an accretion model that attributes the low radiative output to a low mass accretion rate,  rather than a low radiative efficiency. Authors calculated the combined disk-jet steady-state broadband spectrum. A comparison between predicted and observed spectra indicates that M87 may be a maximally spinning black hole accreting at a rate of 
$\sim 10^{-3}M_{sun}yr^{-1}$
 In the resent report made by \cite[Younes et al (2019)]{Younes19} is presented the analysis of simultaneous XMM-Newton+NuSTAR observations of two low-luminosity Active Galactic Nuclei (LLAGN), NGC 3998 and NGC 4579. In the paper were calculated parameters of  X-ray emmision of these LLAGNs, and were discussed these results in the context of hard X-ray emission from bright AGN, other LLAGN, and hot accretion flow models. 

The goal of our study was to obtain XMM spectra for LLAGNs and to search for the common behavior of their spectra. Therefore we performed facial analysis of XMM observations for every source. To find the best spectral fit to the data, several models were tested in this work. For the beginning we fitted every obtained spectra with a simple phabs·powerlaw model that shows that X-ray emission is driven by non-thermal processes. We found that X -ray spectra of only 3 sources from our sample can be fitted with simple phabs·powerlaw model. Another 5 LLAGNs have very complicated spectra and fitting with phabs·powerlaw model doesn`t provide good 
$\chi ^2 $
statistic.

The data extraction was performed with the Science Analysis System (SAS) version 16.0.0, The PN data are selected using event patterns 0–4 and 0–12, respectively. We extracted source events for all observations from a circle that should hold 90\% of X-ray emission of the source and the center should be equidistant from the ring. 

We used archived XMM-Newton observations for 8 from 12 selected LLAGNs. (see Table 1) NGC3642 is not bright in X-ray range and for 3 LLAGNs, namely NGC404 NGC3368 and NGC4203 there is no XMM observations. For sources with more than 5 XMM observations we chose observations distributed in time with the reason that we can observe some changes in their spectral properties, if they exist (see Table 2 ).

 Table \ref{tab1}.

\section{Implications}

{\underline{\it Conclusions}}.We created the sample of 12 LLAGNs for investigation of their X-ray properties.  Redshift z in our sample varies from -0.000140 in M81 to 0.00562 in NGC 4579. It means that all objects are relatively close to us galaxies. From Table 1 we can see that 75\% (9 from 12) of the host galaxies of selected LLAGNs are spiral and only 25\% (3 from 12) are elliptical.
We obtained XMM spectra for LLAGNs and tried to find common behaviors in their spectra. Nevertheless their relatively low brightness (if compare with AGNs), we found that X-ray spectra of only 3 sources from our sample can be fitted with simple phabs·powerlaw model. X-ray spectra of another 5 X-ray sources required models with multiple components and shows complicated features. Obtained model's parameters are:  Photon Index varies from 1.2 in NGC 1052 to 2.90 in NGC 3486; absorption by neutral hydrogen $N_H=2.22 \cdot 10^{20} cm^{-2}$ in NGC 3998 to $N_H=7.5\cdot 10^{22} cm^{-2}$.


\begin{thebibliography}{}

\bibitem[Dan Maoz (2008)]{Maoz2008}
{Dan Maoz} 2008,
\textit{ J. Phys.: Conf. Ser.} 131, 012036

\bibitem[Jolley, E.J.D., Kuncic Z. (2007)]{JolleyKuncic07}
{Jolley, E.J.D., Kuncic, Z.}  2007,
\textit{Ap\&SS} 311, 257

\bibitem[Minfeng Gu \& Xinwu Cao (2009)]{GuCao09}
{Minfeng Gu, Xinwu Cao}  2009,
\textit{MNRAS}, 399, 349



\bibitem[ Roberts at all (2001)]{Roberts2001}
{Roberts T. P., N. J. Schurch, R. S. Warwick}  2001,
\textit{MNRAS},  324,  737 


\bibitem[Younes et al (2019)]{Younes19}
{Younes, George; Ptak, Andrew; Ho, Luis C.; Xie, Fu-Guo; Terasima, Yuichi; Yuan, Feng; Huppenkothen, Daniela; Yukita, Mihoko} 2019,
\textit{ApJ}, 870, 73





\end{thebibliography}
\end{document}